# Enhanced electrocatalytic activity via phase transitions in strongly correlated SrRuO$_3$ thin films

Sang A Lee,[a,b,*] Seokjae Oh,[a,*] Jae-Yeol Hwang,[c,d] Minseok Choi,[e,f] Chulmin Youn,[g] Ji Woong Kim,[h] Seo Hyoung Chang,[i] Sungmin Woo,[a] Jong-Seong Bae,[j] Sungkyun Park,[h] Young-Min Kim,[c,d] Suyoun Lee,[k] Taekjib Choi,[g] Sung Wng Kim,[d] and Woo Seok Choi[a,**]

[a]Department of Physics, Sungkyunkwan University, Suwon 16419, Korea

[b]Institute of Basic Science, Sungkyunkwan University, Suwon 16419, Korea

[c]Center for Integrated Nanostructure Physics, Institute for Basic Science (IBS), Suwon 16419, Korea

[d]Department of Energy Sciences, Sungkyunkwan University, Suwon 16419, Korea

[e]Materials Modeling and Characterization Department, Korea Institute of Materials Science, Changwon 51508, Korea

[f]Department of Physics, Inha University, Incheon 22212, Korea

[g]Hybrid Materials Research Center, Department of Nanotechnology and Advanced Materials Engineering, Sejong University, Seoul 05006, Korea

[h]Department of Physics, Pusan National University, Busan 46241, Korea

[i]Department of Physics, Chung-Ang University, Seoul 06974, Korea

[j]Busan Center, Korea Basic Science Institute, Busan 46742, Korea

[k]Electronic Materials Research Center, Korea Institute of Science and Technology, Seoul 02792, Korea

* These authors contributed equally to this work.

** E-mail : choiws@skku.edu

**Transition metal oxides have been extensively studied and utilized as efficient catalysts. However, the strongly correlated behavior which often results in intriguing emergent phenomena in these materials has been mostly overlooked in understanding the electrochemical activities. Here, we demonstrate a close correlation between the phase transitions and oxygen evolution reaction (OER) in a strongly correlated $SrRuO_3$. By systematically introducing Ru-O vacancies into the single-crystalline $SrRuO_3$ epitaxial thin films, we induced phase transition in crystalline symmetry which resulted in corresponding modification in the electronic structure. The modified electronic structure significantly affect the electrochemical activities, so a 30% decrease in the overpotential for the OER activity was achieved. Our study suggests that a substantial enhancement in the OER activity can be realized even within single material systems, by rational design and engineering of their crystal and electronic structures.**

Transition metal oxides show promising chemical activities that can be applied in solid oxide fuel cells (SOFC), rechargeable batteries, catalytic converters, oxygen-separation membranes, and gas sensors.[1-5] Oxygen evolution reaction (OER, $4OH^- \rightarrow O_2 + 2H_2O + 4e^-$) is one of the most important steps in energy conversion and storage mechanisms, and is the efficiency-limiting process in electrolytic water splitting and metal-air batteries.[6,7] The ultimate goal of OER study is to develop low-cost, highly active, and stable catalysts.[8,9] Recently, perovskite oxides ($ABO_3$), such as $Ba_{0.5}Sr_{0.5}Co_{0.8}Fe_{0.2}O_{3-\delta}$, $Pr_{0.5}Ba_{0.5}CoO_{3-\delta}$, and $LaCoO_3$, have attracted much attention owing to their intrinsically high OER activity.[10-12] More interestingly, properties such as surface oxygen binding energy, number of outer shell electrons in the transition metal ion, electron occupancy of the $e_g$ orbitals, and the proximity of the oxygen $p$-band to the Fermi level, have been proposed as descriptors for OER activity.[10,11,13,14] Such approaches, however, have been mainly tested by comparing systems containing different transition metal elements. Unintentionally, such variations in the identity of the elements therein involve commensurate changes in the atomic structures, valence states, electric resistivities, crystalline surfaces, and the overall and specific electronic structures of the materials. Therefore, approaches based on simplified electronic structure may not apply to distinctive material systems, and more carefully controlled study, for example, one using a single-material system, is necessary to precisely understand the effect of the catalyst's electronic structure on the OER.[15]

In order to probe the link between electronic structure and catalytic activity within a single-material system, we exploit the strongly correlated behavior in complex oxides. In particular, the strong coupling among the degrees of freedom of the $d$-electrons, *i.e.*, charge, spin, orbital, and lattice, in transition metal oxides enables effective tuning of the fundamental physical properties of transition metal oxides.[16,17] In the case of epitaxial thin films, the strong coupling results in a wide spectrum of material properties, which is accessible by tuning epitaxial strain and/or defect concentration. Particularly, elemental defects in perovskite oxides can induce modifications in chemical bonding and hybridization between transition metal $3d$ and oxygen $2p$ orbitals, which affect electronic structure significantly. By utilizing these characteristics, we are able to selectively tailor the electronic structure within a single transition metal oxide system.

$SrRuO_3$ (SRO) is a strongly correlated metal that exhibits good OER activity, although it is not very stable in electrochemical environments.[9] Bulk SRO crystals have an orthorhombic structure with lattice parameters of $a_o$ = 5.567, $b_o$ = 5.530, and $c_o$ = 7.845 Å, which can also be represented as a pseudocubic structure with a lattice parameter of $a_{pc}$ = 3.926 Å.[18] Epitaxial SRO thin films exhibit a structural phase transition induced by temperature, epitaxial strain, thickness, and/or vacancy concentration, which makes them suitable for the study of the relationship between fundamental physical properties and OER activities.[19-22] In particular, a structural phase transition to a higher symmetry lattice, *e.g.*, an orthorhombic to tetragonal transition, is observed with increasing entropy (temperature and/or defect concentration) of the system.[19,23]

In this communication, we demonstrate the close correlation between the crystalline symmetry, electronic structure, and OER activity, within the framework of strongly correlated transition metal (Ru 4*d*) and oxygen electronic states. In particular, an increase (decrease) in the unoccupied Ru 4*d* $e_g$ (occupied O 2*p*) electronic state accompanies orthorhombic to tetragonal structural transition in epitaxial SRO thin films, which results in a substantial (> 30%) decrease in the overpotential for the OER.

In order to induce a phase transition in the crystal structure, we fabricated epitaxial SRO thin films on a single-crystalline $SrTiO_3$ (STO) (001) substrate at different oxygen partial pressure, $P(O_2)$, using pulsed laser epitaxy. Fig. 1 shows the crystal structure of our SRO thin films. The thicknesses of the samples were fixed at ~30 nm (Fig. S1, ESI†). The X-ray diffraction (XRD) $\theta$-$2\theta$ scans in Fig. 1(a) represent a monotonic increase of the pseudocubic *c*-axis lattice parameter ($c_{pc}$) with decreasing $P(O_2)$. The lattice expansion originates from the increased Ru and O vacancies with decreasing $P(O_2)$, as will be discussed later in more detail. Despite the introduction of elemental vacancies, all the SRO thin films are coherently strained to the STO substrate, as exemplified in the reciprocal space map around the (103) STO Bragg reflection, shown in Fig. 1(b). All the thin films show good crystallinity, as indicated by the peaks in the rocking curve scans, with full-width-at-half-maximum (FWHM) values of < 0.02° (Fig. S2, ESI†).

We observed an orthorhombic to tetragonal structural phase transition with a decrease in $P(O_2)$ for our fully strained epitaxial SRO thin films. In particular, XRD off-axis $\theta$-$2\theta$ scans show a clear distinction between the films grown at $P(O_2) \geq 3 \times 10^{-2}$ and $\leq 1 \times 10^{-2}$ Torr (Fig. S3, ESI†).[23-26] Fig. 1(e) summarizes the calculated lattice parameters for the epitaxial SRO thin films grown at varying $P(O_2)$ obtained by the above XRD analyses (Table S1, ESI†). An abrupt change in the orthorhombic lattice parameters is observed across $P_c(O_2) = \sim 2 \times 10^{-2}$ Torr, which is the critical $P(O_2)$ for the structural phase transition. Consequently, crystal structures for the epitaxial SRO thin films are schematically shown in Figs. 1(c) and 1(d), *i.e.*, a distorted orthorhombic structure (Fig. 1(d)) for the films grown at $P(O_2) > P_c(O_2)$ and a tetragonal structure (Fig. 1(c)) for the films grown at $P(O_2) < P_c(O_2)$, within the preserved perovskite framework. (The lattice parameters are redefined according to the crystal symmetry for each structure.)

The $P(O_2)$-dependent orthorhombic-to-tetragonal phase transition is driven by the increased elemental vacancy concentrations within the SRO epitaxial thin films. We performed X-ray photoemission spectroscopy (XPS), which confirms a systematic and gradual increase of the Ru and O vacancies with decreasing $P(O_2)$. While it is technically difficult to determine the exact stoichiometry of oxide thin films, we can understand the qualitative trend of the vacancy formation. Figs. 2(a) and 2(b) show Ru 3$d$ with an asymmetric peak shape and O 1$s$ core level photoemission spectra, respectively. The peaks at binding energies of 285.5, 281.4, and 529.0 eV correspond to Ru $3d_{3/2}$, $3d_{5/2}$ (spin-orbit splitting of 4.1 eV), and O 1$s$ orbitals, respectively.[27] From the XPS spectra we clearly observe a systematic decrease in the Ru 3$d$ spin-orbit doublets and an increase in the O vacancy states with decreasing $P(O_2)$.[28] Fig. 2(c) shows the relative atomic concentration of each element in the thin films, obtained from the spectral weight of the deconvoluted peaks. With decreasing $P(O_2)$, a continuous decrease in Ru and O concentrations are observed, along with an increase in O vacancy concentration ($V_O$) and the secondary phase (Sr-O). Interestingly, the elemental vacancy concentrations do not show a drastic increase across $P_c(O_2)$. Hence, the structural phase transition is not induced by abrupt changes in the vacancy concentrations. Instead, despite a small difference, when a critical amount of elemental vacancies is introduced, the energy of sustaining the orthorhombic structure becomes too high and the structure lowers the energy by transforming into a tetragonal structure. Theoretical calculation of the vacancy formation energy further supports the elemental vacancy engineering of the crystal structure. (Table S2, ESI†)

Concomitant with the structural phase transition, we observed a substantial modification in the electronic structure of SRO thin films across $P_c(O_2)$. Fig. 3 shows optical spectroscopic results obtained by spectroscopic ellipsometry, which reflects the electronic structure near the Fermi level. The real part of the optical conductivity spectra, $\sigma_1(\omega)$, (Fig. 3(a)) clearly exhibits evolution of optical transitions with changing $P(O_2)$, especially across the structural phase transition. SRO thin films typically show a Drude absorption at low photon energy, indicating metallic behavior. In addition, four different peaks, labelled as $α$, A, $β$, and B at ~1.7, ~3.3, ~4.1, and ~6.2 eV, are attributed to Ru 4$d$ $t_{2g}$ → $t_{2g}$, O 2$p$ → Ru 4$d$ $t_{2g}$, Ru 4$d$ $t_{2g}$ → $e_g$, and O 2$p$ → Ru 4$d$ $e_g$ optical transitions, respectively.[29,30] In particular, the optical transition peaks A and B directly represent the quantitative hybridization strength between the Ru and O states. In order to extract individual parameters for each optical transition, we used the following Drude-Lorentz analysis,

$$\sigma_1(\omega) = \frac{e^2}{m^*}\frac{n_D \gamma_D}{\omega^2 + \gamma_D^2} + \frac{e^2}{m^*}\sum_j \frac{n_j \gamma_j \omega^2}{(\omega_j^2 - \omega^2)^2 + \gamma_j^2 \omega^2}, \qquad (1)$$

where, $e$, $m^*$, $n_j$, $γ_j$, and $ω_j$ are the electronic charge, effective mass, carrier density, scattering rate, and resonant frequency of the $j$-th oscillator, respectively. The first term describes the coherent Drude response from free charge carriers (subscript $D$), and the second term describes the optical transitions from occupied to unoccupied electronic states. Although the position ($ω_j$) and the width ($γ_j$) of the optical transition peaks do not change significantly, a strong redistribution of the spectral weight ($W_s \equiv \int \sigma_j(\omega) d\omega \propto n_j/\omega_j^2$) is observed with decreasing $P(O_2)$, as shown in Fig. 3(b). In particular, with increasing Ru-O vacancy concentration, $W_{sB}$ and $W_{sβ}$ show an abrupt decrease and increase, respectively, across the structural phase transition. $W_{sβ}$ even surpasses $W_{sA}$ for the tetragonal phase SRO thin films, indicating a strong modification of the electronic structure.

The abrupt change of the electronic structure can be understood by considering the crystal structure as well as the elemental vacancy concentration. First, we note that both A and B are charge transfer transitions involving an occupied O 2$p$ state. Therefore, the transition probabilities ($W_{sA}$ and $W_{sB}$) might decrease as more O vacancies are introduced with decreasing $P(O_2)$. It also indicates that the hybridization strength between the Ru and O states is substantially decreased as the crystal structure changes into a tetragonal structure. In addition, the increase in $W_{sβ}$ can be understood in terms of the Ru-O-Ru bond angle. As the crystal structure undergoes a phase transition from orthorhombic to tetragonal at $P_c(O_2)$, the Ru-O-Ru bond angle approaches 180°, facilitating inter-site $d$-$d$ transitions due to the greater overlap of 4$d$ orbitals. Conversely, $W_{sα}$ shows little, if any, $P(O_2)$ dependence, since the bond angle dependence will be weaker for the $t_{2g}$ orbitals compared to the elongated $e_g$ orbitals that are responsible for the transition $β$. The more abrupt nature of $W_{sB}$ compared to $W_{sA}$ can also be explained by the involvement of the unoccupied $e_g$ orbital.

With the help of density functional theory calculation, we can further confirm that the electronic structure transition is predominantly associated with the crystal symmetry change. Fig. 4 shows the electronic density of states of the orthorhombic and tetragonal SRO. The overall features of the two phases are quite similar, reproducing all the expected optical transitions as indicated by the horizontal arrows in Fig. 4(a). Moreover, essential differences in the predicted electronic structure are clearly observed, as indicated by the vertical arrows in Fig. 4(b). For the tetragonal SRO, the occupied O 2$p$ level (near –2 eV) is suppressed, while the unoccupied Ru 4$d$ $e_g$ level (near 1 eV) is enhanced. In particular, the increase of the unoccupied $e_g$ level, or closing of the energy gap around 1 eV, in the tetragonal SRO is closely related to the octahedral distortion in the orthorhombic structure, as previously discussed. Indeed, it has been shown that octahedral rotation leads to gap opening in 4$d$ transition metal oxides, including $Sr_2RuO_4$.[31,32] It should be further noted that the elemental vacancies do not contribute significantly to the changes in the density of states when compared to the contribution made by the structural transition. Fig. S4 (ESI†) shows the effect of RuO vacancies in determining the electronic structure of the tetragonal SRO thin films, which is relatively weak compared to the effect of the structural transition. (The main changes denoted by the vertical arrows are undisturbed by introducing vacancies.)

The modification in the electronic structure in the epitaxial SRO thin films influences the electrochemical activity significantly. Fig. 5 shows the current density as a function of potential in reference to a reversible hydrogen electrode (RHE) for the SRO thin films in KOH solution. Our cyclic voltammetry data for the orthorhombic SRO thin film ($P(O_2) \geq 3 \times 10^{-2}$ Torr) shows good agreement with a recent study on epitaxial SRO thin films. Chang *et al.* described the current enhancement in the cyclic voltammetry mainly due to the OER activity in SRO, separately confirmed by rotating ring disk electrode (RRDE) experiments on Ru metal.[9] Furthermore, a lower onset potential value (< 1.25 V) has been observed in our tetragonal SRO, which is close to the calculated thermodynamic equilibrium potential value (1.23 V).[14] The extremely low overpotential let us consider partial contribution from transient anodic current based on the oxidation of the SRO thin film itself. However, from cyclic voltammetry at low potential, XRD, XPS, inductively coupled plasma-mass spectroscopy (ICP-MS), and chronoamperometric measurements, we confirmed that both orthorhombic and tetragonal SRO thin films are stable at least up to 1.3 V, which is well above the onset potential of OER (Figs. S8-S12, ESI†). Although more transient behavior due to the dissolution of the material could be observed as we further increased the potential, such behavior did not occur at low potential range.

The OER behavior can be clearly distinguished between the orthorhombic and tetragonal SRO thin films. The cyclic voltammetry curves can be easily categorized into two groups, *i.e.*, the tetragonal SRO thin films exhibit a lower onset potential, indicating a higher OER activity, compared to the orthorhombic SRO thin films. These results clearly demonstrate that the crystal and/or electronic phase transition is responsible for the remarkably enhanced OER activity, on top of the more gradual improvements possibly due to vacancy incorporation in the thin films. In addition, we have observed a reproducible pre-peak in the cyclic voltammetry below the onset of the OER. While its exact origin is not currently understood,[33,34] the pre-peak is specific for the orthorhombic (~1.12 V) and tetragonal (~1.04 V) SRO thin film catalysts, further suggesting a correlation between the electronic structure and catalytic activity. The inset of Fig. 5 shows that the overpotential for the film grown at high $P(O_2)$ ($3 \times 10^{-2}$ Torr) is 0.233 V, while that for the film grown at low $P(O_2)$ ($1 \times 10^{-2}$ Torr) is 0.154 V. Therefore, a 30% reduction in the overpotential is achieved with the concomitant structural and electronic phase transitions in SRO thin films.

We provide further experimental evidence to support that the significant decrease in the overpotential is due to the change in the electronic structure.. First, we provide the electric resistivity as a function of temperature (Fig. S5, ESI†). Interestingly, the resistivity increases with increasing defect concentration (decreasing $P(O_2)$), which is the opposite trend to the increasing current density observed in the alkaline solution. Therefore, the decrease in the overpotential for the tetragonal SRO in KOH solution cannot be attributed to a more conducting sample, and is indeed due to the enhanced chemical activity. Second, the orthorhombic and tetragonal SRO epitaxial thin films show comparable structural and surface quality, as shown in Fig. S2 (ESI†). The rocking curve scan indicates the excellent crystallinity of the samples, and the atomic force microscopy (AFM) topography images show the single unit cell step-and-terrace structure of the substrate, which is well-preserved even after thin-film growth, for the samples grown just above and below $P_c(O_2)$. The rms roughness values obtained from the topographic images are also comparable. The XRD $\theta$-$2\theta$ scans also show prominent thickness fringes for the SRO films grown down to $P(O_2) = 1 \times 10^{-2}$ Torr, which is below $P_c(O_2)$, indicating that the surface structural quality is not the main factor for the enhancement of the OER activity in the tetragonal SRO thin films. Moreover, since we do not observe any abrupt increase in the Ru and O vacancy concentrations (Fig. 2), the modification in the electronic structure the best natural explanation for the enhanced catalytic activity in the tetragonal SRO thin films.

The simplified schematic band diagrams in the inset of Fig. 5 illustrates the essential modification in the electronic structure that should be associated with the change in observed electrochemical activity. We specifically note that the hybridization strength between the Ru and O states within the SRO thin films can substantially influence the OER activity. Indeed, metal-oxygen hybridization in $3d$ transition metal oxides has been recently proposed as an important factor in determining the electrocatalytic activity.[3,35,36] The mechanism of OER in alkaline solution consists of series of elementary electron-transfer steps, the efficiencies of which are determined by the creation and breaking of chemical bonds between the adsorbate molecules such as $O^*$, $OO^*$, $OH^*$, and $OOH^*$ and the surface transition metal ions.[37,38] As the A peak in the optical spectroscopy is a straightforward measure of the hybridization strength, the tetragonal SRO thin film has lower hybridization strength compared to the orthorhombic thin film. The reduced hybridization within the $RuO_6$ octahedra can facilitate the formation of the chemical bonding between the Ru and the adsorbate molecules. In addition, the increase in the unoccupied $e_g$ state might indicate the easier formation of more oxidized (unstable) active Ru site, which could further facilitate the activity.[9]

A recent scanning tunneling microscopy study also underscored the important role of the $RuO_6$ octahedra for understanding water molecule adsorption at the ruthenate surfaces.[39] In particular, the structural flexibility or "rigidity" of the octahedra in perovskite oxides was identified as the key factor influencing molecular adsorption. The exact difference between the orthorhombic and tetragonal SRO surfaces at the molecular level is yet to be discovered, in terms of difference in bonding character and possible surface superstructures. Nevertheless, the effect of structural and electronic modification on the surface chemistry and reactivity in a single-material system has been clearly demonstrated.

**Conclusions**

In conclusions, we have studied the close correlation between the crystalline symmetry, electronic structure, and OER activity of single-crystalline SrRuO$_3$ thin films. A critical oxygen partial pressure of ~2 × 10$^{-2}$ Torr during the pulsed laser epitaxy growth clearly divided the grown epitaxial thin films into two groups, *i.e.*, orthorhombic and tetragonal phase for high and low pressure growth, respectively. The elemental-vacancy-induced phase transition in the crystalline lattice accompanied a change in the electronic structure. The modification of the electronic structure affected the electrocatalytic activity substantially, reducing the overpotential of the OER by more than 30%. Our approach provides an effective framework for discovering an appropriate descriptor for different chemical activities, and can be applied to other strongly correlated transition metal oxide catalysts.

**Acknowledgements**

S.A.L. and S.O. contributed equally to this work. We thank Janice L. Musfeldt, Changyoung Kim, and S. S. Ambrose Seo for valuable discussions. We appreciate S.-J. Park for her help on 2D-RSM measurement at Korea ITS Co. Ltd. This work was supported by Basic Science Research Programs through the National Research Foundation of Korea (NRF) (NRF-2014R1A2A2A01006478, NRF-2016R1A6A3A11934867 (S.A.L.), NRF-2015R1C1A1A02037595 (M.C.), NRF-2015R1C1A1A01053163 (S.H.C.), NRF-2015R1D1A1A01058672 (S.P.)), and NRF-2014R1A1A006405 (T.C.)). S.L. was supported by the Korea Institute of Science and Technology (KIST) through 2E25800. This work was also supported by IBS-R011-D1.

# Figures

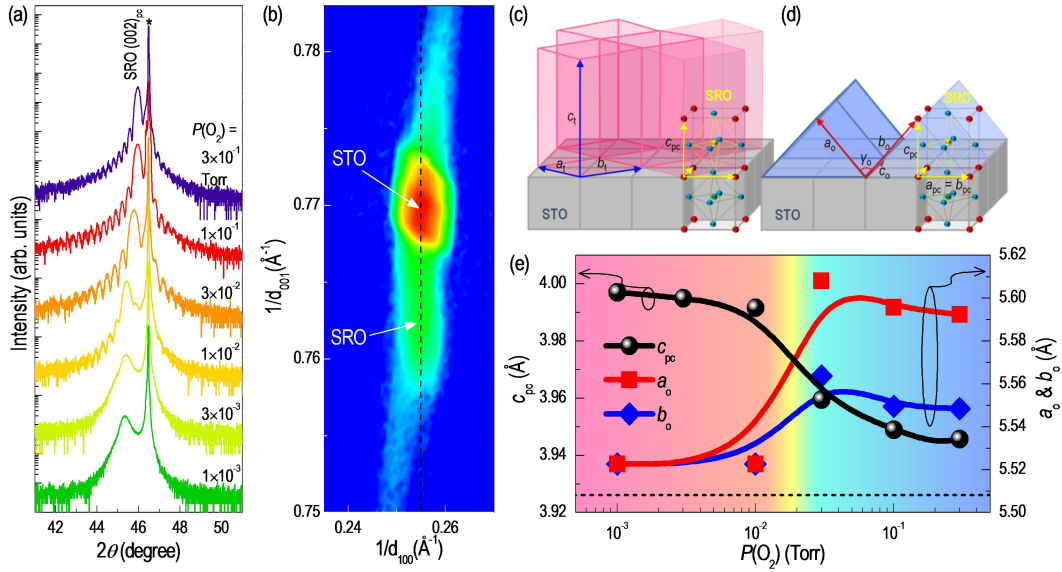

**Fig. 1 | Elemental vacancy induced structural phase transition in epitaxially strained SrRuO$_3$ thin films.** (a) XRD $\theta$-$2\theta$ scans for epitaxial SrRuO$_3$ thin films grown at different $P$(O$_2$) around the (002) Bragg reflections of the SrTiO$_3$ substrates (*). With decreasing $P$(O$_2$), the (002)$_{pc}$ peak of SrRuO$_3$ shifts to a lower angle, indicating an increase of the $c_{pc}$-axis lattice constant. (b) XRD reciprocal space map of the SrRuO$_3$ thin film grown at $P$(O$_2$) = $10^{-1}$ Torr around the (103) Bragg reflection of the SrTiO$_3$ substrate, which shows a coherently strained film without any lattice relaxation. Schematic diagrams of (c) tetragonal and (d) orthorhombic SrRuO$_3$ thin films epitaxially grown on SrTiO$_3$ substrates. The lattice parameters are redefined according to the crystal symmetry for each structure. (e) Evolution of the orthorhombic and pseudocubic lattice constants of epitaxial SrRuO$_3$ thin films as a function of $P$(O$_2$). A clear structural phase transition is observed at $P_c$(O$_2$) = ~2 × $10^{-2}$ Torr. The pseudocubic lattice constant of bulk SrRuO$_3$ ($c_{pc}$ = 3.926 Å, black dashed line) is shown for comparison. Background colors represent the tetragonal (red) and orthorhombic (blue) phase regions.

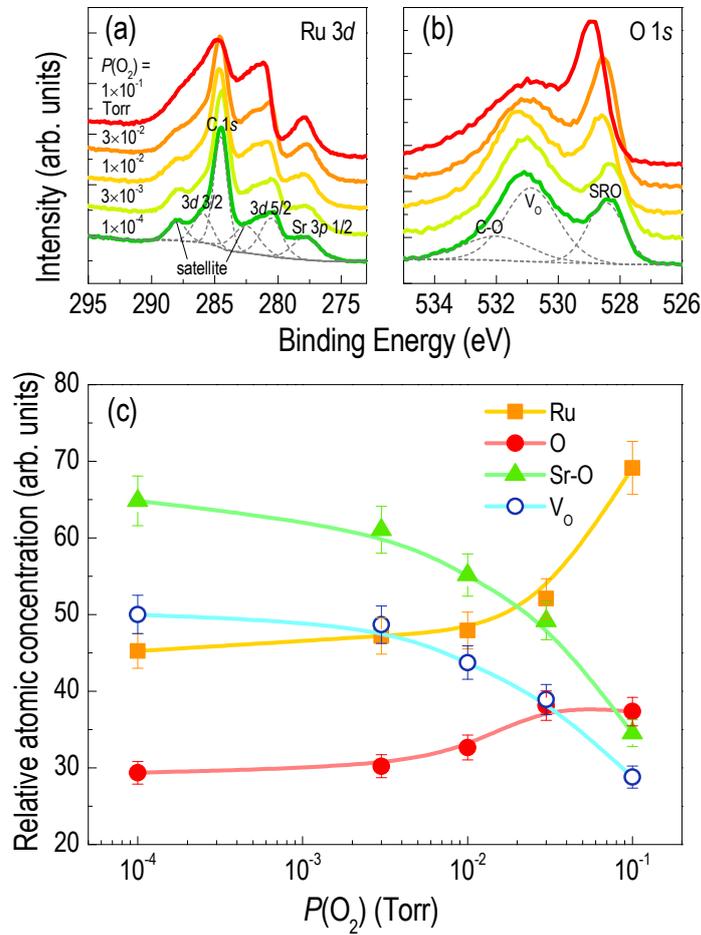

**Fig. 2 | Elemental defect concentration in SrRuO$_3$ thin films by X-ray photoemission spectroscopy.** (a) Ru 3$d$ core-level and (b) O 1$s$ XPS spectra for the SrRuO$_3$ thin films grown at different $P$(O$_2$). The spectra are vertically shifted for clarity. Starting from stoichiometric SrRuO$_3$, an increase of Ru and O vacancies is shown with decreasing $P$(O$_2$). (c) Relative atomic concentration in SrRuO$_3$ thin films, showing a systematic increase of Ru and O vacancies with decreasing $P$(O$_2$). While the crystal structure of the SrRuO$_3$ thin films shows a clear phase transition at $P_c$(O$_2$) = ~2 × 10$^{-2}$ Torr, the Ru and O vacancy concentrations do not show such an abrupt change.

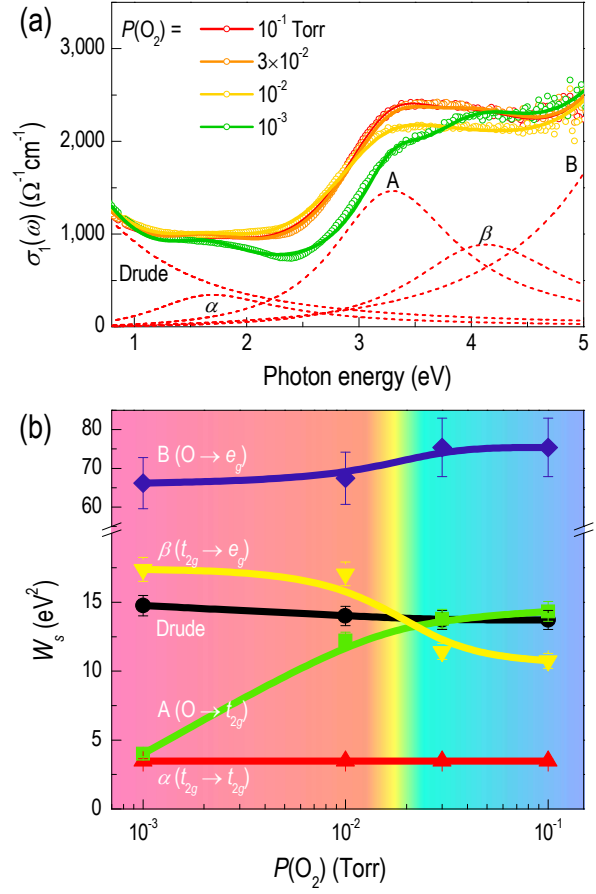

**Fig. 3 | Elemental defect induced electronic structure transition in SrRuO$_3$ thin films.** (a) Real part of the optical conductivity, $\sigma_1(\omega)$, of the SrRuO$_3$ thin films. The symbols denote the experimental spectra and the lines represent the Drude-Lorentz fitting results. The optical transitions at ~1.7, ~3.1, ~4.4, and ~6.2 eV are assigned to peak $\alpha$, A, $\beta$, and B, respectively. The transitions correspond to d-d transition between Ru 4d $t_{2g}$ states ($\alpha$), d-d transition between Ru 4d $t_{2g}$ and $e_g$ states ($\beta$), charge transfer transition between O 2p and Ru 4d $t_{2g}$ states (A), and charge transfer transition between O 2p and Ru 4d $e_g$ states (B). The dashed lines represent the deconvoluted Lorentzian peaks corresponding to each characteristic optical transition. (b) Spectral weight ($W_s$) evolution of the Lorentz oscillators as a function of $P(O_2)$. With decreasing $P(O_2)$, $W_{sA}$ and $W_{sB}$ decrease, while $W_{s\beta}$ increases. In particular, $W_s$ for most transitions shows an abrupt change at $P_c(O_2) = $ ~$2 \times 10^{-2}$ Torr, indicating a transition in the electronic structure.

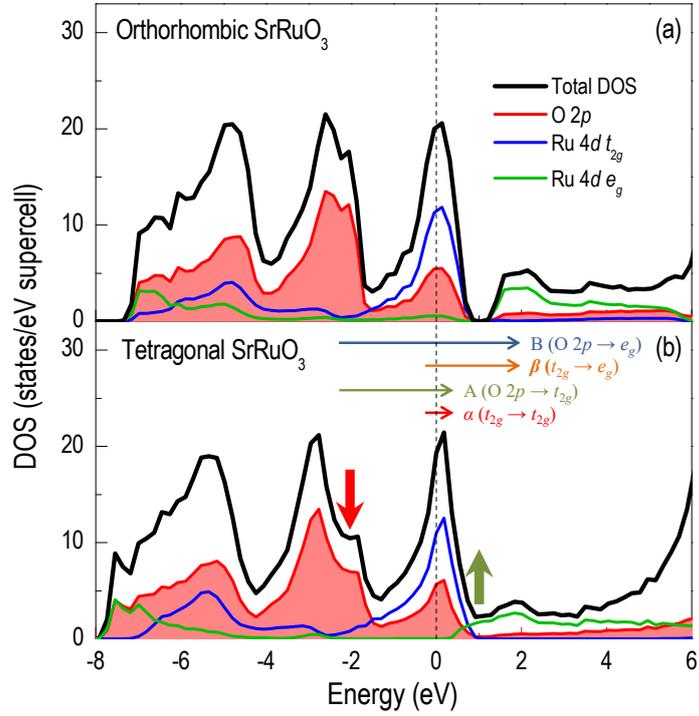

**Fig. 4 | Density of states (DOS) of (a) orthorhombic and (b) tetragonal SrRuO$_3$.** The density of states indicates a decrease in the occupied O 2$p$ band and an enhancement of the unoccupied Ru 4$d$ $e_g$ band as the structure changes from orthorhombic to tetragonal. The horizontal arrows indicate the optical transitions observed in Fig. 3. The change in the theoretical electronic structure owing to the phase transition coincides with the change in the experimental optical spectra.

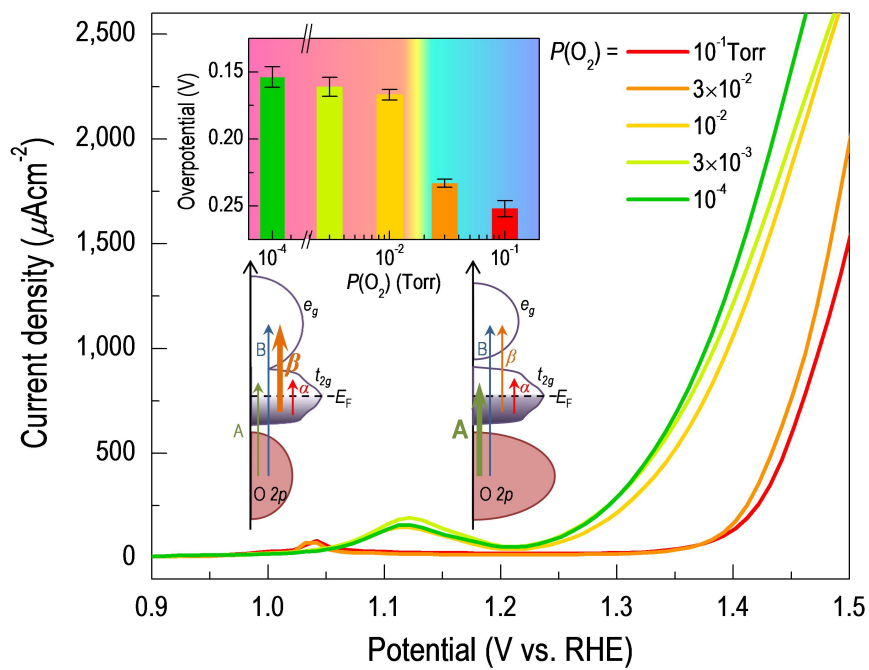

**Fig. 5 | Catalytic OER effect of SrRuO₃ thin films with elemental vacancies.** Current-potential curves during the first potential sweep for the OER on SRO thin film electrodes in a 1 M KOH electrolyte. As the $P(O_2)$ decreases from $10^{-1}$ to $10^{-4}$ Torr, the overpotential also decreases. The insets show the overpotential at current density of 1000 $\mu A\ cm^{-2}$ as a function of $P(O_2)$. Schematic band diagrams are also shown for the tetragonal (larger Ru $4d\ e_g$ state) and the orthorhombic (larger O $2p$ state) structures of epitaxial SRO thin films.

**Electronic Supplementary Information**

**Enhanced electrocatalytic activity via phase transitions in strongly correlated SrRuO$_3$ thin films**

Sang A Lee,[a,b*] Seokjae Oh,[a*] Jae-Yeol Hwang,[c] Minseok Choi,[d,e] Chulmin Youn,[f] Ji Woong Kim,[g] Seo Hoyung Chang,[h] Sungmin Woo,[a] Jong-Seong Bae,[i] Sungkyun Park,[g] Young-Min Kim,[c,j] Suyoun Lee,[k] Taekjib Choi,[f] Sung Wng Kim,[j] and Woo Seok Choi[a]

[a.] Department of Physics, Sungkyunkwan University, Suwon 16419, Korea

[b.] Institute of Basic Science, Sungkyunkwan University, Suwon 16419, Korea

[c.] Center for Integrated Nanostructure Physics, Institute for Basic Science (IBS) Sungkyunkwan University, Suwon 16419, Korea

[d.] Materials Modeling and Characterization Department, Korea Institute of Materials Science, Changwon 51508, Korea

[e.] Department of Physics, Inha University, Incheon 22212, Korea

[f.] Hybrid Materials Research Center, Department of Nanotechnology and Advanced Materials Engineering, Sejong University, Seoul 05006, Korea

[g.] Hybrid Materials Research Center, Department of Nanotechnology and Advanced Materials Engineering, Sejong University, Seoul 05006, Korea

[h.] Department of Physics, Chung-Ang University, Seoul 06974, Korea

[i.] Busan Center, Korea Basic Science Institute, Busan 46742, Korea

[j.] Department of Energy Sciences, Sungkyunkwan University, Suwon 16419, Korea

[k.] Electronic Materials Research Center, Korea Institute of Science and Technology, Seoul 02792, Korea

**Experimental detail**

**Thin film growth and structural characterization**

High-quality epitaxial SrRuO$_3$ thin films were grown on TiO$_2$-terminated SrTiO$_3$ (for optical experiments) and 0.5 wt% Nb-doped SrTiO$_3$ (for Oxygen evolution reaction (OER) analyses) substrates using pulsed laser epitaxy (PLE). A KrF excimer laser ($\lambda$ = 248 nm; Lightmachinery, IPEX 864) with a fluence of ~1.5 J cm$^{-2}$ and repetition rate of 2 Hz was used for the ablation of a sintered target. The substrate temperature was fixed at 700 °C with varying $P(O_2)$ ranging from 3 × 10$^{-1}$ to 1 × 10$^{-4}$ Torr.[1,2] The vertical distance between the target and the substrate was fixed to 65 mm. Crystalline structures of the thin films were determined using high-resolution X-ray diffraction (XRD, Rigaku SmartLab). The thicknesses of films were fixed at ~30 nm, as measured by X-ray reflectometry (XRR) (Fig. S1(a)). Figs. S1(b) and S2 show XRD $\theta$-$2\theta$ scans and rocking curves of the thin films grown at different $P(O_2)$. The results indicate well-oriented single-crystalline thin films without any secondary phases and good crystalline quality (FWHM ≤ 0.02°). An atomic force microscope (AFM, Park Systems NX10) with a Si probe tip (Budget sensors ContAl-G) was used to confirm the surface topography (inset of Fig. S2). Root-mean-square (RMS) roughnesses of 0.687 and 0.821 nm were obtained for the SrRuO$_3$ thin films grown above ($P(O_2)$ = 3 × 10$^{-2}$ Torr) and below ($P(O_2)$ = 1 × 10$^{-2}$ Torr) $P_c(O_2)$, respectively, indicating atomically flat surfaces with preserved step-and-terrace atomic structures of the substrate for both samples. Fig. S3 shows {204} SrTiO$_3$ Bragg reflections with $\varphi$ angles of 0, 90, 180, and 270°. The parameter of the SrRuO$_3$ orthorhombic unit cell parameter of $a \neq b \neq c$ and $\alpha \approx \beta \approx \gamma \approx 90°$ with additional distortion can be calculated to the pseudocubic unit cell ($a_{pc}$ = $a_{STO}$, $b_{pc}$ = $a_{STO}$, $c_{pc}$) through the following relationships,[3,4]

$$a_{\text{pc}} = \frac{c_o}{2},$$

$$b_{\text{pc}} = \frac{\sqrt{a_o^2 + b_o^2 + 2a_o^2 b_o^2 \cos\gamma_o}}{2},$$

$$c_{\text{pc}} = \sqrt{\frac{a_o^2 + b_o^2 - 2b_{pc}^2}{2}}.$$

Table S1 shows the calculated lattice parameters of SrRuO$_3$ thin films on SrTiO$_3$ substrates grown at various $P(O_2)$.

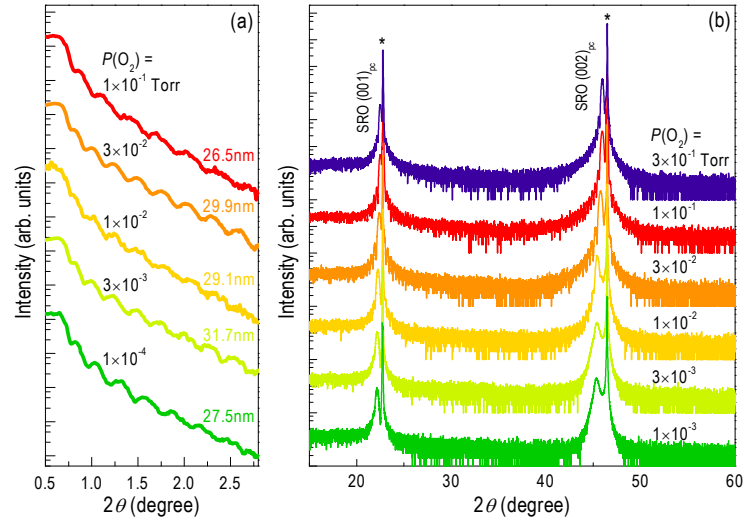

**Fig. S1.** XRR and XRD $\theta$–$2\theta$ scans for SrRuO$_3$ thin films on SrTiO$_3$ substrates grown at different $P(O_2)$. (a) XRR results show that SrRuO$_3$ thin films have thickness of ~30nm. (b) XRD results indicate single-crystalline SrRuO$_3$ thin films are grown over a wide range of $P(O_2)$ without any secondary phases.

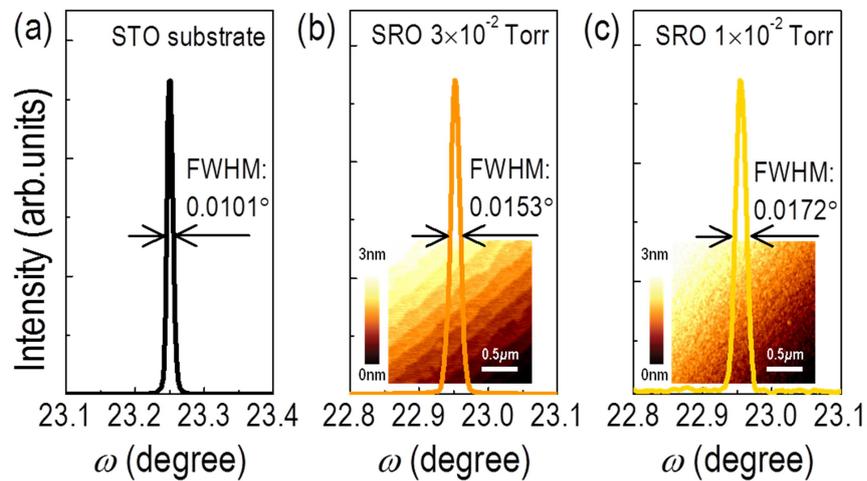

**Fig. S2.** Rocking curve scans and atomic force microscopy images of epitaxial SrRuO$_3$ thin films. Rocking curve scans of (a) SrTiO$_3$ substrate, SrRuO$_3$ thin film with (b) orthorhombic and (c) tetragonal structures. Both the AFM images in the insets show the step-and-terrace structure.

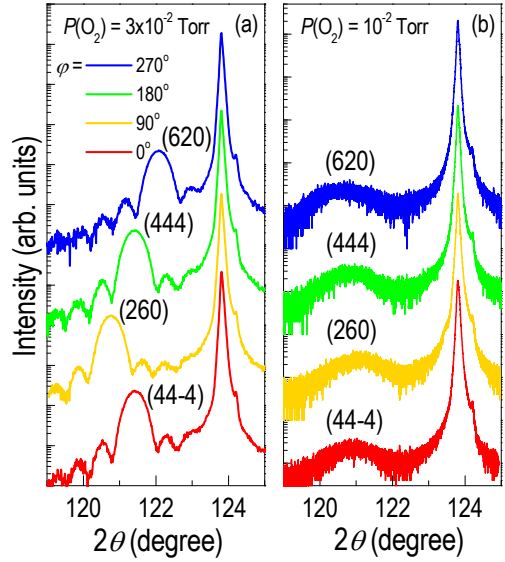

**Fig. S3.** Off-axis X-ray diffraction for the orthorhombic and tetragonal SrRuO$_3$ thin films. (20$L$)$_{pc}$ XRD reflections of SrRuO$_3$ thin films grown at $P(O_2)$ = (a) $3 \times 10^{-2}$ and (b) $1 \times 10^{-2}$ Torr, around the SrTiO$_3$ (204) Bragg reflections with configuration of $\varphi$ = 0, 90, 180, and 270°.

| | $a$ (Å) | $b$ (Å) | $c$ (Å) | $\gamma$ (°) | orthorhombic distortion ($a/b$) |
|---|---|---|---|---|---|
| Bulk[5] | 5.567 | 5.530 | 7.845 | 90 | 1.00669 |
| $3 \times 10^{-1}$ Torr | 5.592 | 5.548 | 7.810 | 89.020 | 1.00794 |
| $1 \times 10^{-1}$ Torr | 5.596 | 5.549 | 7.810 | 88.976 | 1.00847 |
| $3 \times 10^{-2}$ Torr | 5.608 | 5.564 | 7.810 | 88.709 | 1.00791 |
| $1 \times 10^{-2}$ Torr | 5.523 | 5.523 | 7.983 | 90 | 1 |
| $1 \times 10^{-3}$ Torr | 5.523 | 5.523 | 7.986 | 90 | 1 |

**Table S1.** Lattice parameters of SrRuO$_3$ thin films grown at various $P(O_2)$.

**X-ray photoemission spectroscopy**

The chemical structure was studied at room temperature using XPS (Theta Probe, Thermo) with a monochromated Al-$K\alpha$ X-ray source ($h\nu$ = 1486.6 eV). The step size was 0.1 eV at a pass energy of 50.0 eV with a 400 $\mu$m spot size. All the peak positions were calibrated using the C 1s photoemission signal (284.5 eV). In order to analyze the spectra in detail, we deconvoluted the peaks using a mixed Gaussian-Lorentzian function.[6]

**Ellipsometry**

The optical properties of the SrRuO$_3$ thin films were investigated using spectroscopic ellipsometers (VASE and M-2000, J. A. Woollam Co.) at room temperature. The optical spectra were obtained between 0.74 and 5.5 eV for incident angles of 70 and 75°. A two-layer model (SrRuO$_3$ thin film on SrTiO$_3$ substrate) was sufficient for obtaining physically reasonable spectroscopic dielectric functions of SrRuO$_3$ that reproduced the literature spectrum.

**Theoretical calculation**

The calculations were performed using the projector augmented-wave method[7] and the generalized gradient approximation exchange-correlation functional[8] with a Hubbard-$U$ correction, as implemented in the Vienna ab initio simulation package.[9] The electronic wave functions were described using a planewave basis set with an energy cutoff of 400 eV. A rotationally invariant +$U$ method[10] was applied to the Ru 4$d$ ($U_{eff}$ = 2.1 eV) orbitals, the value for which was used in previous literature.[11,12] The calculations for vacancy defects in SrRuO$_3$ were performed using 160-atom supercells. The wavefunctions were expanded in a plane-wave basis set with an energy cutoff of 400 eV, and integrations over the Brillouin zone were carried out using the 4×4×2 $k$-point mesh. The atomic coordinates were relaxed until the force acting on each atom was reduced to less than 0.05 eV Å$^{-1}$. The formation energy of each vacancy was evaluated by the equation in a reference,[13] by considering the secondary phases of SrO and RuO$_2$. As an estimate for the sample growth, we set the oxygen chemical potentials ($\mu_O$) to –1.39 and –1.58 eV, corresponding to $P(O_2)$ values of 10$^{-1}$ and 10$^{-3}$ Torr, respectively at a temperature of 700 °C.

Theoretical calculation of the vacancy formation energy supports the elemental vacancy engineering of the crystal structure. As shown in Table S2, vacancies are likely to form in the tetragonal SrRuO$_3$, and the RuO vacancy may predominantly prevail due to it having the lowest formation energy, which is consistent with the spectroscopic observations. For an orthorhombic structure, the formation of the vacancies would be relatively suppressed since the formation energy is much higher compared to that of tetragonal SrRuO$_3$.

| Vacancy type | Tetragonal structure (eV) | Orthorhombic structure (eV) |
|---|---|---|
| Ru | 0.30 | 0.54 |
| O | -0.19 | 1.55 |
| RuO | -0.36 | 1.09 |

**Table S2.** Formation energy of different types of vacancies in SrRuO$_3$ thin films with distinctive crystal structures. Theoretical calculation of formation energy (in eV) suggests that O and RuO vacancies spontaneously form in the tetragonal SrRuO$_3$ thin films, while the formation energy in the orthorhombic structure is positive for all the types of defects studied.

In order to estimate the contribution of defects compared to that of the structural phase transition, we calculated the density of states (DOS) of stoichiometric orthorhombic SrRuO$_3$ (Fig. S4(a)), stoichiometric tetragonal SrRuO$_3$ (Fig. S4(b)), and RuO defect-induced tetragonal Sr$_{32}$Ru$_{31}$O$_{95}$ (Fig. S4(c)). The main modifications in the electronic structure indicated by optical spectroscopy are represented by red and green arrows: The occupied O 2$p$ level decreases (red) and the unoccupied Ru 4$d$ $e_g$ level increases (green) when the crystal structure changes from orthorhombic to tetragonal. Interestingly, the DOS of the stoichiometric (Fig. S4(b)) and the RuO deficient (Fig. S4(c)) tetragonal SrRuO$_3$ have the same features in terms of the main electronic structure, indicating that the structural phase transition is a crucial factor for the electronic structure modification in the SrRuO$_3$ grown at low $P$(O$_2$). In other words, elemental vacancies do not sufficiently contribute to changes in the electronic structure.

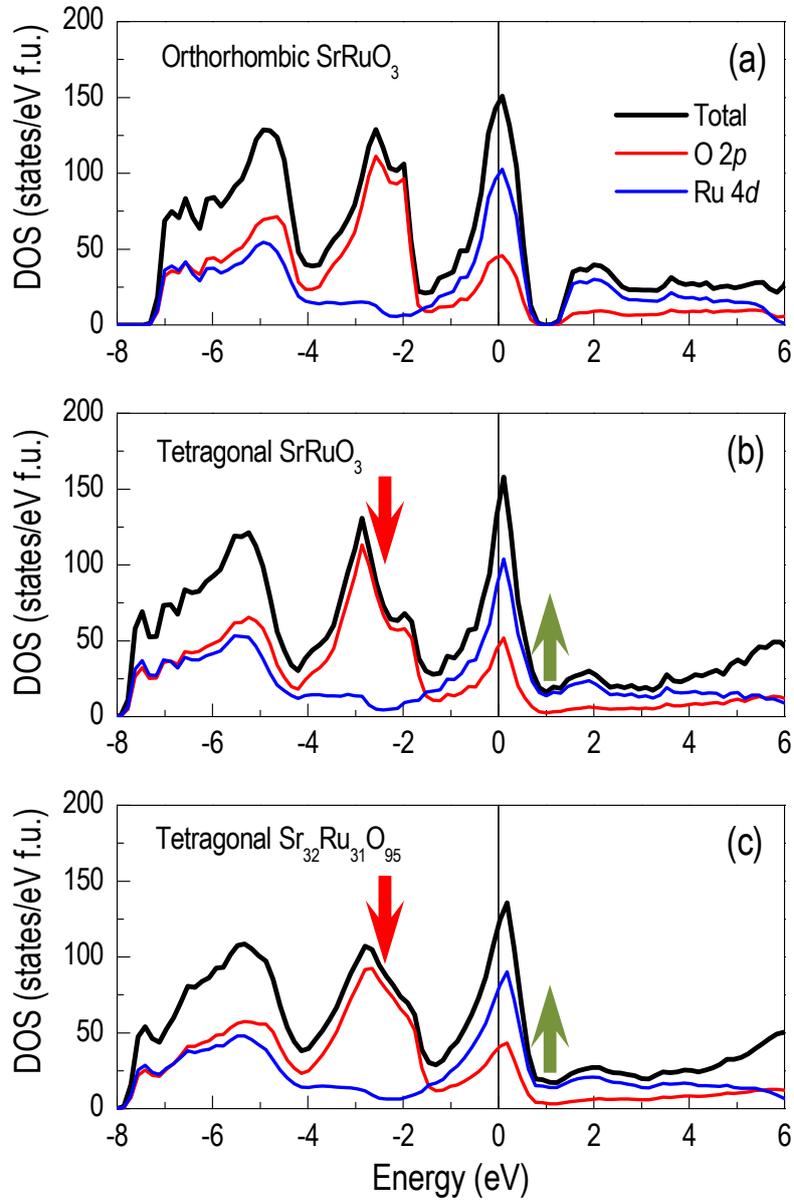

**Fig. S4.** Theoretical calculations of the density of states for the SrRuO$_3$ thin films. (a) Orthorhombic stoichiometric SrRuO$_3$, (b) tetragonal stoichiometric SrRuO$_3$, and (c) tetragonal RuO deficient Sr$_{32}$Ru$_{31}$O$_{95}$ thin films are shown for comparison.

**Resistivity measurements**

Resistivity as a function of temperature, $\rho(T)$, was measured using a low-temperature closed-cycle refrigerator (ARS-4HW, Advance Research Systems). The measurements were performed from 300 to 20 K, using the Van der Pauw method with In electrodes and Au wires.

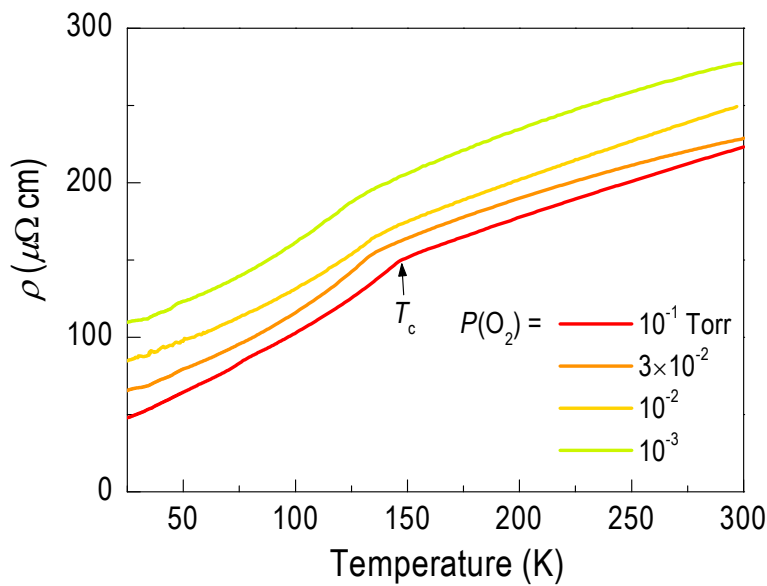

**Fig. S5.** Changes in the electronic properties of SrRuO$_3$ thin films. Resistivity as a function of temperature ($\rho(T)$) is shown for the SrRuO$_3$ thin films deposited at various $P(O_2)$. With decreasing $P(O_2)$, overall $\rho(T)$ increases which could be attributed to vacancy scattering.

**Electrochemical measurements**

A Vertex potentiostat (Ivium Technologies) was used for electrochemical measurements using the three-electrode method. Ag/AgCl (reference electrode), Pt mesh (counter electrode), and high-quality epitaxial SrRuO$_3$ thin films (working electrode) were used as the electrodes. The SrRuO$_3$ films were contacted with copper wires using silver epoxy (TED PELLA, INC.) and there was no noticeable contact resistance between samples and wires. The sample, except for the SrRuO$_3$ film surface, was sealed by coating PMMA, to avoid direct exposure of the conducting substrate to the electrolye. We carefully soaked the SrRuO$_3$ thin films into the electrolyte below the contact region and the area of the working electrode was 4.5 × 4.5 mm$^2$. The schematic diagram of the sample for electrochemical measuremetns is shown in Fig. S6. For the electrolyte, we used 1 M

KOH prepared with deionized water, and the same amount of the surface areas of each sample was carefully soaked into the KOH solution, without submerging the contact region. The sweep rate for cyclic voltammetry was 5-50 mV s$^{-1}$. The trends of the OER activity between the orthorhombic and tetragonal SrRuO$_3$ thin films were the same, independent of the sweep rate. Note that the potential is shown vs. a reversible hydrogen electrode (RHE).

In order to confirm that the large increase in the current level in cyclic voltammetry originates indeed from OER, we performed several additional experiments. First, as shown in Fig. S7, we confirmed that the Nb:SrTiO$_3$ substrates do not affect the OER activity of SrRuO$_3$ thin films. Second, we measured several cyclic voltammetry curves at different potential sweep range to confirm the stability. Fig. S8 shows the reversible current-potential curves in SrRuO$_3$ thin films at lower potential range (initial stability test). These measurements have been conducted on all of our SrRuO$_3$ thin films to confirm the sample are stable at low potential ($\leq$ 1.3 V), which is well above the onset potential of OER (~1.25 V). The data of Fig. 5 in the main text have been taken after these measurements. Based on this result, we believe that the low onset potential in tetragonal SrRuO$_3$ is due to the electrocatalytic reaction, and not coming from the materials oxidation. Third, Fig. S9(a) again shows reversible cyclic voltammetry sweep at low potential ($\leq$ 1.3 V). These samples have been further tested using XRD to confirm that there is no change in the thickness or crystal structure of the thin films after several cycles of the electrochemical experiments. X-ray photoemission spectroscopy (XPS) shown in Fig. S10 also suggests that our samples are stable at low potential ($\leq$ 1.3 V). We further performed the inductively coupled plasma-mass spectroscopy (ICP-MS) and we did not observe any noticeable Sr and Ru dissolution in KOH solution after the stability test up to 1.3 V for several cycles (Fig. S11). Indeed, we observed only small amount (~2 ppb) of Sr in the solution which was comparable to the reference solution and did not observe any Ru in the test solution (up to 1.3 V), for both orthorhombic and tetragonal SrRuO$_3$ thin films. We also show the solution after applying potential up to 1.7 V for comparison, where the elements (both Sr and Ru) are actually dissolved into the solution. Fig. S12 shows the chronoamperometric measurements as compared with cyclic voltammetry. While the SrRuO$_3$ thin film is known to dissolve in alkaline solution at a large applied voltage (> 1.5 V), current density well above the onset potential (up to ~1.4 V) is maintained, indicating consistent OER behavior.

Fig. S13 shows the Tafel plots for all the SrRuO$_3$ thin films with slope values of (i) $1 \times 10^{-1}$ Torr: 77.7 mV dec$^{-1}$, (ii) $3 \times 10^{-2}$ Torr: 66.2 mV dec$^{-1}$, (iii) $1 \times 10^{-2}$ Torr: 107.4 mV dec$^{-1}$, (iv) $3 \times 10^{-3}$ Torr: 108.5 mV dec$^{-1}$, and (v) $1 \times 10^{-4}$ Torr: 106.1 mV dec$^{-1}$, respectively. Again, a clear distinction is demonstrated, indicating the difference in the kinetics of the OER between the SrRuO$_3$ thin films with different crystalline structures (orthorhombic and tetragonal).

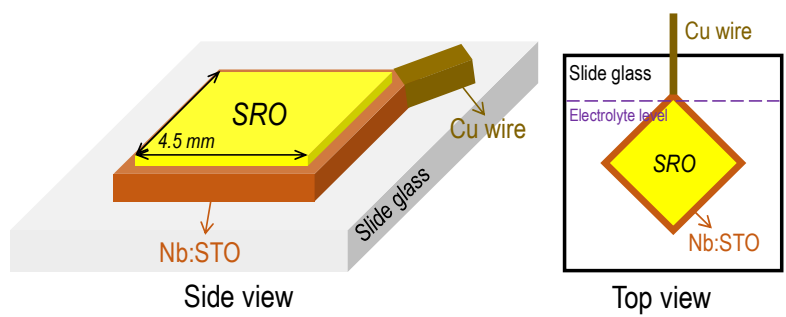

**Fig. S6.** Schematic diagram of SrRuO$_3$ thin film sample for the electrochemical measurements.

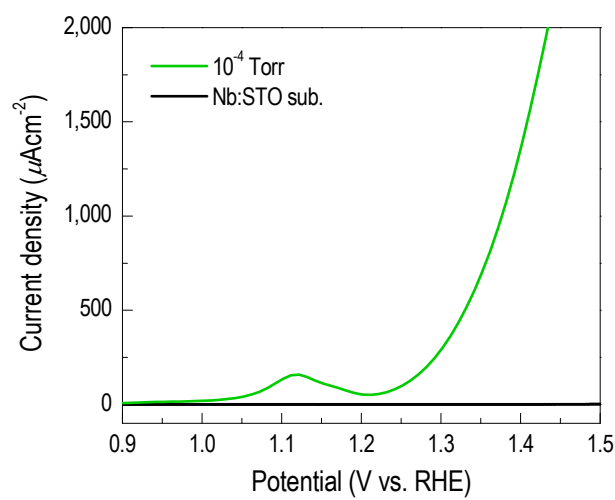

**Fig. S7.** Current-potential curves of SrRuO$_3$ thin film and Nb:SrTiO$_3$ substrate. Nb:SrTiO$_3$ substrate does not show any current in the whole potential range, which allows us to eliminate the contribution of the substrate to the observed OER activity in the SrRuO$_3$ thin films.

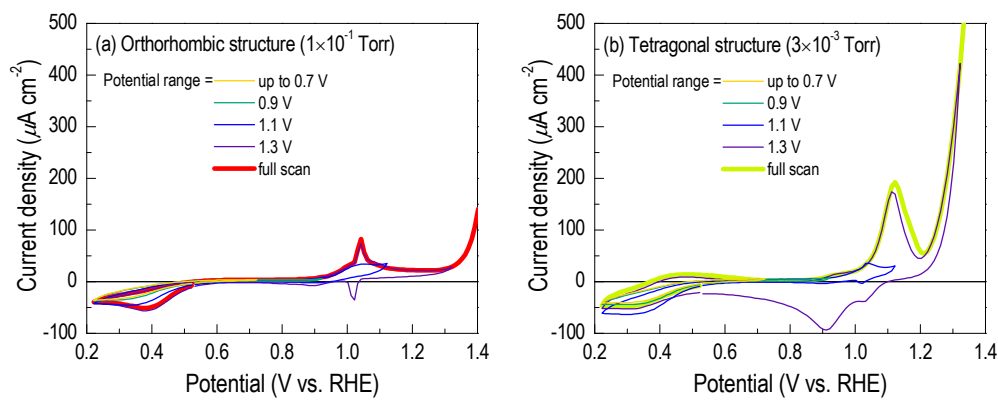

**Fig. S8.** The current-potential cycle in (a) orthorhombic and (b) tetragonal epitaxial SrRuO$_3$ thin films at lower potential range (initial stabilization test). The Figure shows the data with a scan rate of 50 mV s$^{-1}$, but the same systematic trend is observed for the same measurements with lower scan rates (5-50 mV s$^{-1}$).

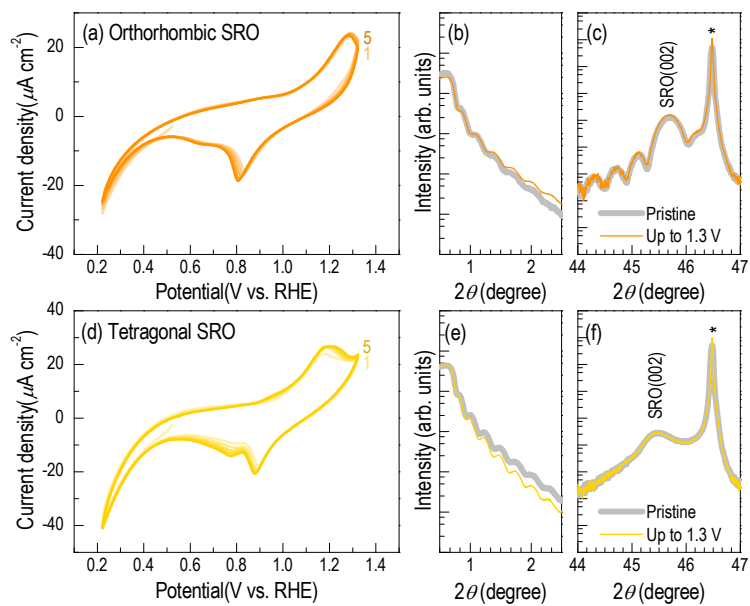

**Fig. S9.** Reversible cyclic voltammetry results at low potential range (< 1.3 V) of (a) orthorhombic and (d) tetragonal SrRuO$_3$ thin films. (b,c,e,f) XRD result for the SrRuO$_3$ sample before and after the initial stability test. Gray lines indicate pristine SrRuO$_3$ thin film and color lines indicate the same sample after the initial stability test, respectively.

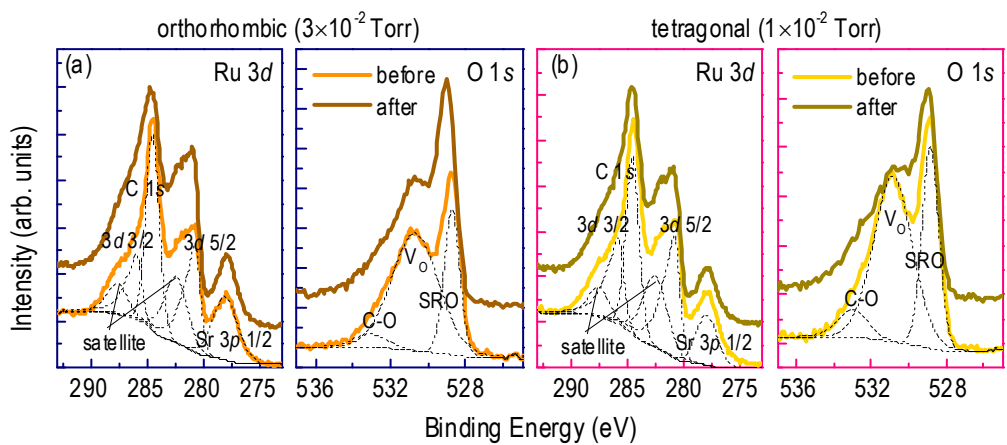

**Fig. S10.** X-ray photoemission spectroscopy for (a) orthorhombic and (d) tetragonal SrRuO$_3$ thin films before and after the initial stability test, respectively.

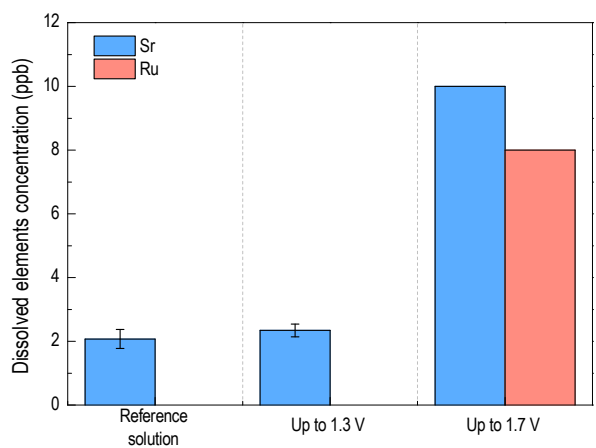

**Fig. S11.** Contents of dissolved elements in the KOH solution obtained by ICP-MS. Our results show negligible amount of dissolution of the sample both orthorhombic and tetragonal structure when the potential is applied up to 1.3 V, which is clearly higher than the onset potential of OER.

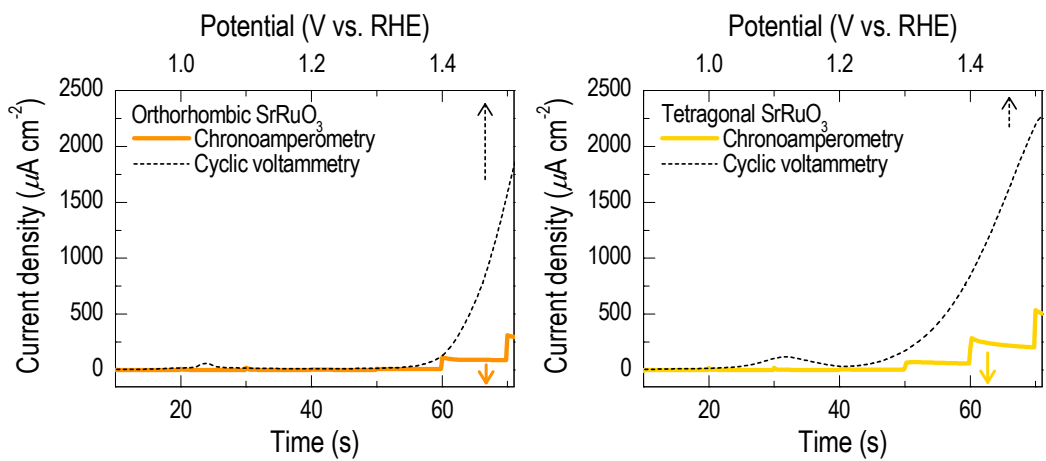

**Figure S12**. Chronoamperometric measurements on the orthorhombic and tetragonal SrRuO$_3$ epitaxial thin films grown at 3 × 10$^{-2}$ and 1 × 10$^{-2}$ Torr, respectively, in dark.

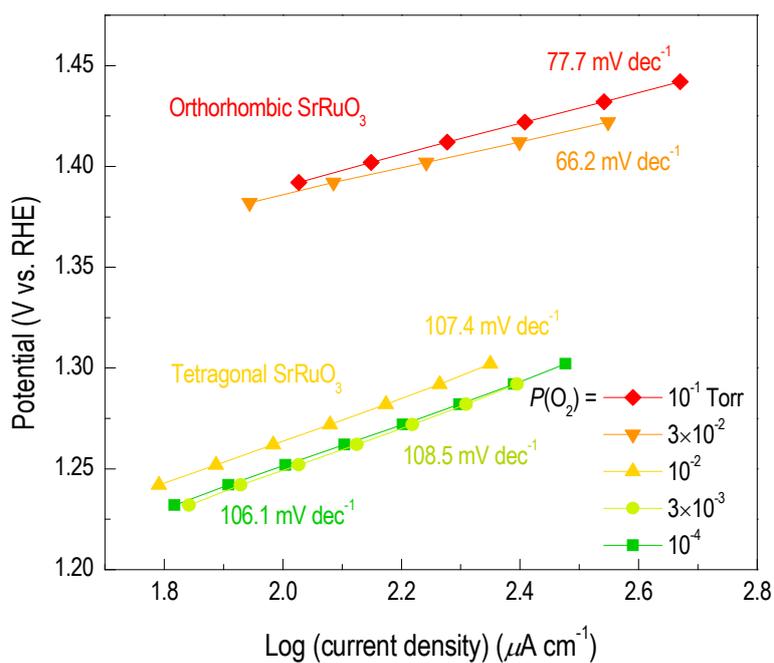

**Fig. S13.** Tafel plots of epitaxial SrRuO$_3$ thin films grown at different $P$(O$_2$). Tafel plots show the tetragonal SrRuO$_3$ films have smaller overpotential than orthorhombic SrRuO$_3$ films.